\documentclass[10pt]{article}

\usepackage{amsmath}

\begin{document}


\begin{center}
{\Large {\bf Dynamical maps and measurements}}
\end{center}

\hfil\break

\begin{center}
{\bf Aik-meng Kuah\footnote{kuah@physics.utexas.edu}, E.C.G. Sudarshan \\}
{\it Department of Physics \\
University of Texas at Austin \\
Austin, Texas 78712-1081}\\
\end{center}

\begin{center}
September 12, 2002
\end{center}

\begin{abstract}

We show how a set of POVMs, expressed as a set of $\mu$ linear maps, can be performed with a unitary transformation followed by a von-Neumann measurement with an ancillary system of no more than $\mu N^2$ dimensions.  This result shows that all generalized linear transformations and measurements on density matrices can be performed by unitary transformations and von-Neumann measurements by coupling a suitably large ancillary system.

\end{abstract}

\hfil\break
\hfil\break

\pagebreak


\section{Introduction}

Dynamical maps (also known as superoperators) describe generalized linear transformations that can be performed on density matrices.  It has been shown that dynamical maps that preserve the trace of density matrix can be performed by an unitary transformation by adding an ancillary system of a given size ~\cite{Sudarshan85}.

In the related problem of measurement, generalized measurements would be described by a set of linear maps, where each individual map acting on the density matrix would not preserve its trace, much like a projection acting on a Hilbert space ray would not preserve its norm.  It has been shown that if the maps can be described by one-dimensional operations, then it can always be performed as a projection in a larger system (ie. with a suitable ancillary system).

In this paper, we show that any set of non-trace preserving maps can be performed by making unitary transformations and von-Neumann (projective) measurements in a larger system.  This result shows that all linear transformations and measurements on density matrices can always be performed by unitary transformations and von-Neumann measurements, by coupling an ancillary system of appropriate size to the original system.

We will begin by re-deriving the result that a dynamical map can be performed by a unitary transformation in a larger system, and we will show that this proof can be easily extended to a set of maps.

\section{A dynamical map}

A dynamical map is linear and maps valid density matrices to valid density matrices:

\begin{eqnarray}
\Lambda: \rho \rightarrow \Lambda_{rr',ss'} \rho_{r',s'} = \rho'_{r,s}	\\
\text{Hermiticity}\quad \rho'_{r,s} = {\rho'_{s,r}}^* \Rightarrow 
\Lambda_{rr',ss'} = {\Lambda_{ss',rr'}}^*	\\
\text{Preserves trace}\quad Tr[\rho'] = Tr[\rho] \Rightarrow \sum_r 
\Lambda_{rr',rs'} = \delta_{r',s'}
\end{eqnarray}

The map should also be completely positive to preserve the positivity of the density matrix.

A map acting on a $N \times N$ density matrix can be written as a $N^2 \times N^2$ hermitian matrix, and therefore it has a canonical decomposition with $\nu \leq N^2$ eigen-operators and real eigenvalues (see ~\cite{Sudarshan61},~\cite{Choi72},~\cite{Kraus83}):

\begin{equation}
\Lambda: \rho \rightarrow \sum_\alpha \lambda_\alpha L_\alpha \rho 
L_\alpha^\dagger
\end{equation}

The condition that the map preserves the trace requires:

\begin{equation}
\sum_\alpha \lambda_\alpha L_\alpha^\dagger L_\alpha = I
\label{norm}
\end{equation}

Let us consider a transformation $U$ acting on a larger system, which consists of the original system coupled with an ancillary system of $\nu$ dimensions.  This larger Hilbert space is spanned by a basis $\{ |r^A\rangle|\alpha^B\rangle; 0 \geq r \geq N-1, 0 \geq \alpha \geq \nu-1 \}$.  Let $U$ be given by:

\begin{equation}
U: |r'^A\rangle|0^B\rangle \rightarrow \sum_{r \alpha} \sqrt{\lambda_\alpha} 
\left[L_\alpha\right]_{rr'} |r^A\rangle|\alpha^B\rangle
\end{equation}

The superscript $A$ labels the original system, and the superscript $B$ labels the ancillary system.  At this point, we have not yet specified how $U$ transforms the states $|r'^A\rangle|\alpha^B\rangle$ for $\alpha \neq 0$, however we notice that $U$ preserves the orthornormality between states 
$|r'^A\rangle|0^B\rangle$ and $|s'^A\rangle|0^B\rangle$ if:

\begin{equation}
\sum_{r\alpha} \lambda_\alpha \left[L_\alpha^\dagger\right]_{s'r} 
\left[L_\alpha\right]_{rr'} = \delta_{r',s'}
\end{equation}

This is nothing more than the condition that the map preserves the trace, equation~\ref{norm}.  So $U$ can be easily "completed" into a unitary transformation by simply making sure that the states $\{|r'^A\rangle|\alpha^B\rangle; \alpha \neq 0\}$ are unitarily transformed into the space orthorgonal to the space spanned by $\{ U |r'^A\rangle|0^B\rangle \}$.  Exactly how $U$ transforms those states is unimportant, so many choices of $U$ exist.

Now if we take our original density matrix $\rho$ and couple it to an ancillary system of dimensions $\nu$, initially in the state $|0^B\rangle\langle0^B|$, the unitary transformation $U$ acting on this overall state gives:

\begin{equation}
\sum_{r s \alpha \beta} \sqrt{\lambda_\alpha} \sqrt{\lambda_\beta}
\left[L_\alpha\right]_{rr'} \rho_{r's'} \left[L_\beta^\dagger\right]_{s's}
| r^A \rangle\langle s^A | \otimes | \alpha^B \rangle\langle \beta^B |
\end{equation}

Tracing over the ancillary system $B$ will give us the desired transformation $\Lambda \rho$.  Therefore, we see that any trace-preserving map can be thought of as the contraction of a unitary transformation acting on a larger system, by suitably coupling an ancillary system of dimension $\nu \leq N^2$ to the original system.

\section{Sets of maps}

A generalized measurement can be described by a set of maps $\{ \Lambda^{(i)} \}$, where the probability of the ith outcome is $Tr [ \Lambda^{(i)} \rho ]$ and the state of the system if the ith outcome is obtained (post-selection) is $\Lambda^{(i)} \rho / Tr [ \Lambda^{(i)} \rho ]$.  Individually, the maps $\Lambda^{(i)}$ would not preserve the trace of density matrix, but if the measurement is complete (ie. total probability is 1) then the overall map $\sum_i \Lambda^{(i)}$ will preserve the trace of the density matrix.

Let us decompose the maps into their eigen-operators and eigenvalues:

\begin{equation}
\Lambda^{(i)} : \rho \rightarrow \sum_\alpha {\lambda^{(i)}}_\alpha 
{L^{(i)}}_\alpha \rho {L^{(i)}}_\alpha^\dagger
\end{equation}

The condition that the trace is preserved gives:

\begin{equation}
\sum_{i \alpha} {\lambda^{(i)}}_\alpha {L^{(i)}}_\alpha^\dagger 
{L^{(i)}}_\alpha = I
\label{sum}
\end{equation}

Let us define a transformation $V$ in a larger space:

\begin{equation}
V: |r'^A\rangle|(0,0)^B\rangle \rightarrow \sum_{r (i,\alpha)} 
\sqrt{\lambda^{(i)}_\alpha} \left[L^{(i)}_\alpha\right]_{rr'} 
|r^A\rangle|(i,\alpha)^B\rangle 
\end{equation}

The ancillary system here is now labelled with 2 indices $(i,\alpha)$, and its size is given by the number of operators $\L^{(i)}_\alpha$.  This size is bounded by $\mu N^2$, where $\mu$ is the number of maps $\Lambda^{(i)}$, since each map $ \Lambda^{(i)}$ has at most $N^2$ eigen-operators.

As in the previous section, we have not defined the transformation on a complete set of states, but it preserves the orthornormality between the states $|r'^A\rangle|(0,0)^B\rangle$ and $|s'^A\rangle|(0,0)^B\rangle $:

\begin{equation}
\left(V|r'^A\rangle|(0,0)^B\rangle \right)^\dagger 
\left(V|s'^A\rangle|(0,0)^B\rangle \right) =
\sum_{r i \alpha} \lambda^{(i)}_\alpha
\left[L^{(i)}_\alpha\right]^*_{rr'} \left[L^{(i)}_\alpha\right]_{rs'} = 
\delta_{r' s'}
\end{equation}

The last equality is obtained by applying equation~\ref{sum}.  We can therefore complete $V$ into a unitary transformation as in the previous section.  Now $V$ acting on the initial state $\rho \otimes |(0,0)^B\rangle\langle(0,0)^B|$ will give:

\begin{equation}
\sum_{i j \alpha \beta}
\sqrt{\lambda^{(i)}_\alpha} \sqrt{\lambda^{(j)}_\beta}
\left[L^{(i)}_\alpha\right]_{rr'}
\rho_{r' s'}
\left[L^{(j)}_\beta\right]^*_{ss'}
| r^A\rangle\langle s^A| \otimes | (i,\alpha)^B\rangle\langle (j,\beta)^B | 
\end{equation}

If we make a von-Neumann measurement on the ancillary system, given by the projections $\{ \left(\sum_\alpha |(i,\alpha)^B\rangle\langle (i,\alpha)^B|\right) \}$, the ith outcome would give $\Lambda^{(i)} \rho$.

\section{Non trace preserving maps and sets of maps}

We had required in the last section that the combination of all the maps, $\sum_i \Lambda^{(i)}$, preserves the trace of the quantum state, so it would represent a complete measurement.  If $\sum_i \Lambda^{(i)}$ does not preserve the trace, we need only add a map $I - \sum_i \Lambda^{(i)}$ into the set of maps to be performed.  Any outcome due to this extra map is discarded, this has the effect of reducing the trace of the quantum state.

In the case where we want to perform a dynamical map $\Lambda$ that does not preserve the trace, we can perform the measurement procedure as outlined in the last section on the set of maps $\{ \Lambda, I-\Lambda\}$, and discard the outcome if $I-\Lambda$ is obtained.

\section{Conclusions}

We have shown that all linear measurements and transformations on density matrices can be accomplished by "traditional" von-Neumann measurements and unitary transformations, when performed in a suitably large system.

\section{Acknowledgments}

We would like to thank Anil Shaji and Todd Tilma, for their help and insightful discussions.


\end{document}